\documentstyle[aps,prl,epsf,rotate]{revtex}
\epsfverbosetrue
\begin{document}

%\twocolumn[\hsize\textwidth\columnwidth\hsize\csname @twocolumnfalse\endcsname

\begin{title}
{\bf Conductance Anomalies for Normal-Insulator-Superconductor Contacts\\
%  \\
}
\end{title}
\author{ Sungkit Yip }
\address{
%\begin{instit}
Department of Physics \& Astronomy, Northwestern University,
 Evanston, IL 60208 \footnote{electronic mail: yip@sneffels.phys.nwu.edu}
%\end{intstit}
}
\date{ \today}
\maketitle
\begin{abstract}
{\small
This paper considers the conductance through a dirty junction
between a normal conductor and a superconductor with an 
insulating barrier.  It is shown that for large barrier 
resistances there is a relative enhancement of the conductance near
zero voltage, whereas for low barrier resistance this anomaly
appears at finite voltage.
}
\end{abstract}

%\vskip1pc]

%\narrowtext
%\textwidth 5.5in
%\textheight 9in

\bigskip

Recent experiments on contacts between semiconductors and superconductors
have generated a lot of interest 
(see e.g. \cite{klapwijk94,kastalsky91,nguyen,nitta,bakker} and
references therein).  Besides possible technical application
of these devices, one is interested in understanding how electrical 
conductance below the superconducting gap arises. 
While there is little doubt that the basic process is 
Andreev reflection \cite{andreev}, the role of impurities is
only being appreciated gradually recently.  The interplay between
impurity scattering and the potential scattering at the Schottky 
barrier between the semiconductor and superconductor is believed
\cite{zaitsev90,volkov92a,volkov92b,volkov93,volkov94,wees92,takane93,marmos93}
to be the physics behind the zero bias anomaly (ZBA) observed in
many experiments (e.g. \cite{klapwijk94,kastalsky91,nguyen,nitta,bakker})

To understand the transport in these normal-insulator-superconductor
structures two major theoretical tools have been employed.
In Ref \cite{takane93,marmos93}
 the conductance is expressed
in terms of the transmission matrices across the structure. 
Unfortunately quite often the necessary averages are not available
from random matrix theory and thus numerical calculations to obtain the transmission matrices have to be done for
 a large number of "samples" and then  averages over the samples 
taken.  An alternative approach\cite{zaitsev90,volkov92a,volkov92b,volkov93}
is to use the quasiclassical green's function where the impurity average
has already been made.  However, in this approach the impurity scattering
is only included in the self-consistent Born approximation.  
Moreover these references  confined themselves to the cases of 
high potential barriers (and often low energies) 
so that analytical progress can be made, although 
the basic equations are applicable  more generally
 so long as we are in the dirty limit.
In this paper we shall solve these equations numerically in order
to obtain a better understanding of the physics as well as a detail
comparison with the results from the scattering matrix approaches.
\cite{takane93,marmos93}

We shall then consider a (quasi-one-dimensional) normal conductor 
("wire") situated
between a normal ($x=0$) and a superconducting ($x=L$)  reservoirs.
The reservoirs are assumed to be at equilibrium at potential
$V$ and $0$ respectively.  
%Thus strictly speaking the wire has to
%be part of a constriction (?).
A potential barrier is located at $x_b$ where $ 0 \le x_b \le L$.
We shall confine ourselves to the dirty limit where all the energy
scales are smaller than the impurity scattering rate.  
We also require that the mean free path $l$ to be much less than $L$. 
 (Note that  experiments cited 
above may not satisfy this condition). 
We shall assume that there is no pairing interaction in our conducting "wire"
and we shall consider only ordinary and magnetic impurity scatterings.
To study the transport through the system we shall solve the Usadel
equation
\begin{equation}
[\epsilon \tau_3 - \check \sigma_{spin}, \check g] + 
 {D \over \pi} \partial_\mu (\check g  \partial_\mu \check g) = 0
\label{usadel}
\end{equation}
together with the normalization condition
$\check g ^2 = - \pi^2 \check 1 \ $ 
governing the angular averaged Green's function 
\begin{displaymath}
\check g = 
\left (
\begin{array}{cc}
\hat g^R & \hat g^K \\
0 & \hat g^A
\end{array}
\right )\
\end{displaymath}
and $\hat g^{R,A,K}$ are the retarded, advanced, and Keldysh 
 martix Green's
functions  in particle-hole and spin space.
Here
$\epsilon$ is the energy and $ \hat \sigma_{spin}^{R,A,K} =
{\gamma \over 2 \pi} \hat \tau_3 \hat g^{R,A,K} \hat \tau_3 $ is
the self-energy for magnetic scattering with the scattering rate $\gamma$.
The equation connecting $\check g$ on the two sides of an interface
has been derived by Kuprianov and Lukichev \cite{kuprianov88},
\begin{equation}
\bar \sigma S \check g  \partial_\mu \check g
  = {G^n_b\over 2}  [ \check g (x_{b-}), \check g (x_{b+}) ]
\label{bcg}
\end{equation}
$\bar \sigma $ is the conductivity of the wire,
$S$ the area, and $ G_b^n$ is the conductance
through of the barrier in
the {\bf normal} state.
For our purpose $\hat g^R$ can be chosen to have only $\tau_3$ and
$\sigma_2 \tau_1$ components:
$\hat g^R = - i \pi ( g_3 \tau_3 - f_1 \sigma_2 \tau_1)$.
$\hat g^A$ is related to $\hat g^R$ by 
$\hat g^A = \tau_3 ( \hat g^R) ^{\dagger} \tau_3$.
Since $ g_3^2 + f_1^2 = 1$ from the normalization condition, we can
write 
$g_3 = {\rm cos} \theta$ and 
$f_1 = {\rm sin} \theta $
where $\theta \equiv \theta_r + \theta_i $ is in general complex. From
eq(\ref{usadel})  $\theta$ satisfies
\begin{equation}
2 i ( \epsilon + i \gamma {\rm cos} \theta ) { \rm sin} \theta
  + 
D \partial_x^2 \theta =0
\label{bulkt}
\end{equation}
The boundary condition eq(\ref{bcg}) reduces to
\begin{equation}
( \bar \sigma S ) \partial_x \theta = 
{  G^n_b} {\rm sin} ([\theta]_b) 
\label{bct}
\end{equation}
here $[\theta]_b$ is the discontinuity of $\theta$ across the interface,
i.e., $[\theta]_b = \theta ( x_{b+}) - \theta (x_{b-}) $.
 $\theta$ at $x=0$ and $x=L$ are assumed to attain their equilibrium
values of the reservoirs.  Thus at $x=0$ $\theta_r = \theta_i = 0 $
and at $x=L$, $\theta_r = { \pi \over 2}, \theta_i = { 1 \over 2} 
{\rm ln}  { \Delta + \epsilon \over \Delta -  \epsilon  }$,
i.e., ${\rm tanh} \theta_i = {\epsilon \over \Delta }$ for 
$\epsilon < \Delta $. 
( and  $\theta_r = 0, \theta_i = { 1 \over 2} 
{\rm ln}  { \epsilon + \Delta \over   \epsilon -\Delta }$,
 for 
$\epsilon > \Delta $ ).

We can define a distribution function $\hat h$ by
$\hat g^K = \hat g^R \hat h - \hat h \hat g^A$ where
 $\hat h$ can be
chosen diagonal $\hat h = h_0 \hat \tau_0 + h_3 \hat \tau_3$.
For our purposes it is sufficient to solve for $h_3$, which satisfies
(from eq(\ref{usadel}))
\begin{equation}
\partial_x ( {\rm cosh}^2 \theta_i \partial_x h_3 ) = 0
\label{bulkh}
\end{equation}
in the bulk together with the boundary condition
(from eq(\ref{bcg}))
\begin{equation}
( \bar \sigma S ) {\rm cosh}^2 \theta_i \partial_x h_3 =  \\
{  G^n_b }{\rm cos} ([\theta_r]_b) 
  {\rm cosh} \theta_i (x_{b+}) {\rm cosh} \theta_i (x_{b-})
   [h_3]_b
\label{bch}
\end{equation}
Here $[h_3]_b $ is the discontinuity of $h_3$ across the barrier.
For a potential $V$ applied to the normal reservoir at $x=0$ and
zero potential at the superconducting reservoir, we have then
$h_3(\epsilon) = { n(\epsilon + eV) - n(\epsilon-eV) \over 2} $ at $x=0$ and
$h_3(\epsilon) = 0 $ at $x=L$.  Here 
$n(\epsilon) = {\rm tanh} { \epsilon \over 2 T} $.
We shall confine ourselves to zero temperture, where $n(\epsilon)$ becomes
a step function.  Thus  $h_3(\epsilon)$ decreases from  $ 1$ at $x=0$
to $0$ at $x=L$ if $ \vert \epsilon \vert < eV$, and 
is identically zero otherwise.   
Since the current 
$I(V) =    { \bar \sigma S \over e} \int_{-\infty}^{\infty}
 {  \ d \epsilon \over 16 \pi^2}  
  {\rm tr} [ \tau_3 ( \hat g^R \partial_x \hat g^K + 
                    \hat g^K \partial_x \hat g^A )] $
 can be written as
$I(V) = - { \bar \sigma S \over e} \int_0^{\infty} \ d \epsilon
  {\rm cosh}^2 \theta_i \partial_x h_3  $,
 the conductance
$G(V) \equiv {dI \over dV}$ at voltage $V$ is given by
\begin{equation}
G(V) = - \bar \sigma S {\rm cosh}^2 \theta_i \partial_x h_3 
\vert_{\epsilon = eV}
\end{equation}
It is obvious from the discussion above and eqs(\ref{bulkh},\ref{bch}) 
that $h_3$ behaves like a voltage
along a wire with ( position dependent) conductivity 
$ \bar \sigma S {\rm cosh}^2 \theta_i $
in series with a resistor with conductance 
\begin{equation}
 G_T(V) = G_b^n {\rm cos} ([\theta_r]_b) 
  {\rm cosh} \theta_i (x_{b+}) {\rm cosh} \theta_i (x_{b-})
   \vert_{\epsilon = e V}
\label{GT}
\end{equation}
and can be easily found.  $h_3$ of course should not be taken literally
as the voltage itself, but the concept of these effective conductances
will help us understand our results.  The conductance through the system
can therefore be written as
\begin{equation}
G(V) = ( R_1(V) + R_T(V) + R_2(V) ) ^{-1}
\end{equation}
where
\begin{equation}
R_1(V) = { 1 \over  \bar \sigma S} \int_0^{x_b} dx {\rm sech}^2 \theta_i (x)
  \vert_{\epsilon = eV}
\label{R1}
\end{equation}
and similarly for $R_2$ except the integration limits are from 
$x_b$ to $L$ and $R_T \equiv (G_T(V))^{-1}$.  
The above equations have essentially been derived in 
ref. \cite{volkov92a,volkov94}

Before describing our results it is helpful to discuss the physical
meanings of $\theta_r$ and $\theta_i$.  The usual density of states
is given by ${\rm Re} g_3 = {\rm cos} \theta_r {\rm cosh} \theta_i$.
Notice however this is not the  quantity that is directly relevant
to the effective conductances through the wire or  the barrier. 
 As seen from  eqs(\ref{R1}) and (\ref{GT}) 
the relevant factors are 
${\rm cosh}^2 \theta_i =  ({\rm Re} g_3)^2 + ({\rm Re} f_1)^2$
 and $ {\rm cos} ([\theta_r]_b) 
  {\rm cosh} \theta_i (x_{b+}) {\rm cosh} \theta_i (x_{b-}) =
 ({\rm Re} g_3(x_{b-})) ({\rm Re} g_3(x_{b+})) + 
  ({\rm Re} f_1(x_{b-})) ({\rm Re} f_1(x_{b+}))$ 
 respectively.
Notice also under a unitary transformation between the particle and
hole components of opposite spin (i.e., a generalized Bogoliubov
transformation), $\hat g^R$ transforms into
$\hat g'^R = e^{ i {\chi \over 2} \sigma_2 \tau_2} \hat g^R 
                e^{ - i {\chi \over 2} \sigma_2 \tau_2} $ where
$\chi$ is a real number.  If one parametrizes $\hat g'^R$ by $\theta'$
as before then $\theta_r' = \theta_r + \chi$ and $\theta_i' = \theta_i$.
It is  thus useful to think of $\theta_r$ as the angle of a vector
in an imaginary space which I shall refer to as spectral space
(see also \cite{nazarov94}), and ${\rm cosh} \theta_i$ as its magnitude.
The equation of motion for $\theta_r$ and $\theta_i$ is
given by eq(\ref{bulkt}) and (\ref{bct}).
$ {\rm Re} g_3$ and $  {\rm Re} f_1 $
are then the two ($g$- and $f$-) components of the spectral vector
or the  density of states vector (DOSV).  The effective conductance of
the conductor at a given point and voltage is thus enhanced 
by the square of the DOSV (at that point and voltage), 
while the conductance through a barrier is modified by the factor
corresponding to the scalar product of the spectral vector on the two sides.
Both these factors are invariants under the particle-hole transformation.
Conductance through a barrier is thus possible only when 
the spectral vectors on the two sides have finite projection onto
each other.  For a direct contact between a normal metal 
($\theta_r=0$) and a superconductor ($\theta_i = 0$ for $\epsilon<\Delta$)
there is thus no subgap conductance.  In the NIS structure as in here the finite
resistivity of the wire allows $\theta_r \ne 0$ at $x_{b-}$
and hence a finite barrier conductance. 

We shall concentrate on the subgap conductance $eV << \Delta$.  For 
a better display of the $\theta$ parameters in Fig. \ref{fig:theta} below, 
I have chosen to present the results for $ x_b = 0.92 L$.  The
results for the total resistances presented below are virtually independent
of $x_b$ so long as $x_b \approx L$.  From the above equations
it is convenient to define a dimensionless parameter $g_b$ as
the ratio of the normal state conductances through the barrier
to that of the wire, i.e. $ g_b = G^n_b / G^n_{wire}$, where
$G^n_{wire} = \bar \sigma S/L $.  We shall also express all energies
in  terms of  the natural scale $E_D \equiv D/L^2$. 
The gap $\Delta$ is chosen to be $ 100 E_D$.

\begin{figure}
\centerline{ 
	\epsfysize=0.4\textwidth \rotate[l]{
	\epsfbox{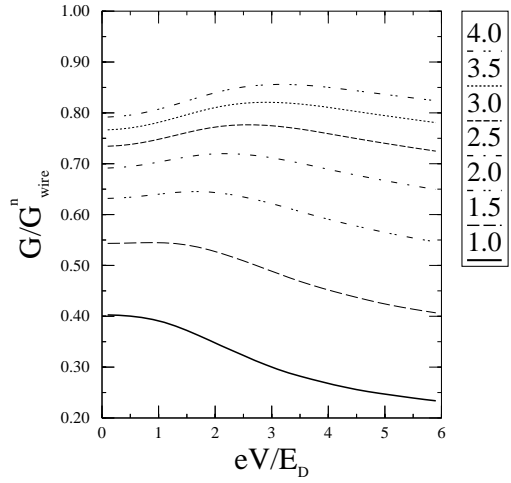} }
}
\caption{
The (normalized) conductance versus voltage for $g_b$'s listed in the legend.  $\gamma=0$ . }
\label{fig:gb}
\end{figure}

\begin{figure}
\centerline{ 
	\epsfysize=0.4\textwidth \rotate[l]{
	\epsfbox{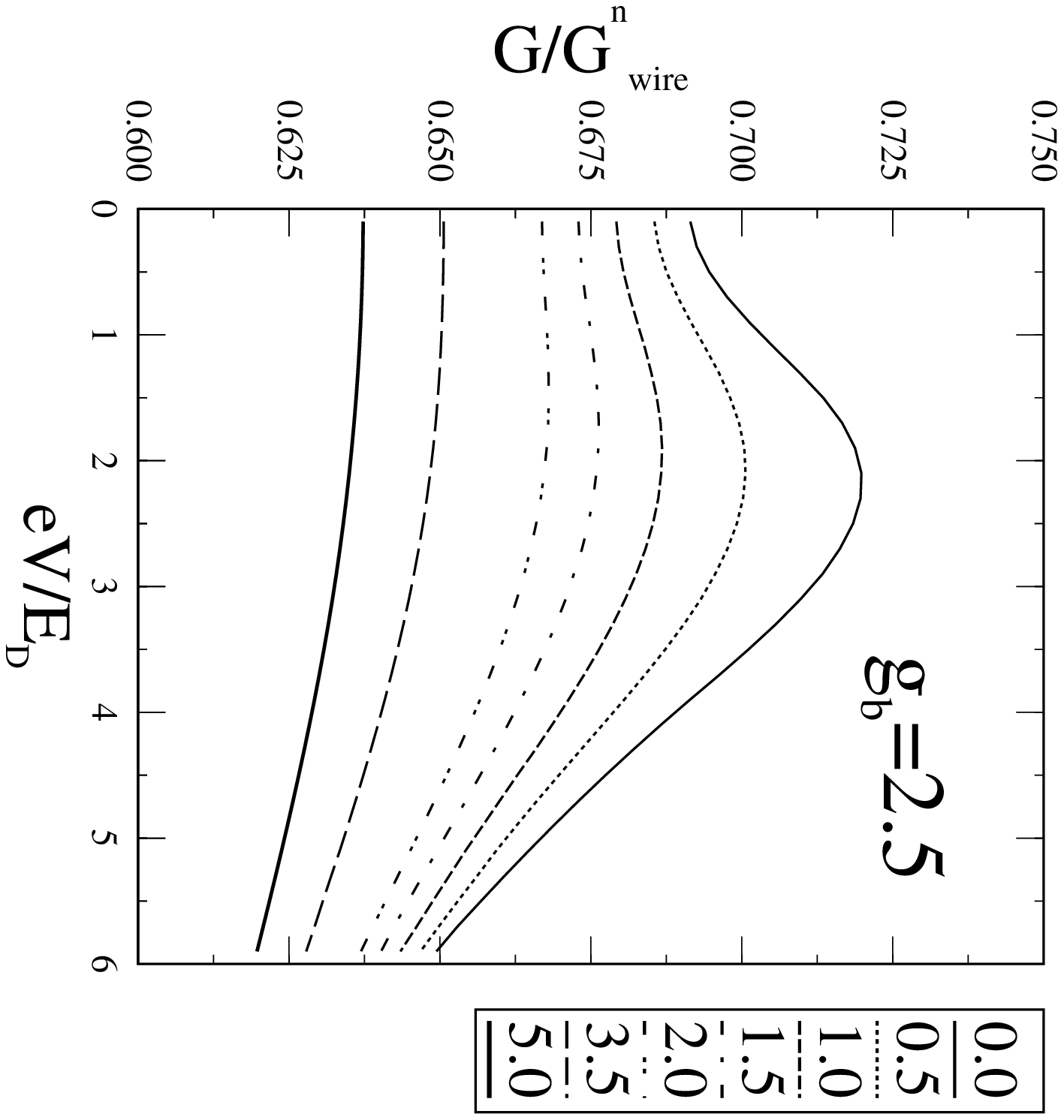} }
}
\caption{
The (normalized) conductance versus voltage at $g_b = 2.5$ for
the values of $\gamma$  listed in the legend. }
\label{fig:gb2.5}
\end{figure}

Fig. \ref{fig:gb} shows the conductance of our 
NIS system for intermediate values
of $g_b$.  For $g_b { < \atop \sim} 1.0$ 
we obtain the an anomaly at zero bias (ZBA). 
 At higher $g_b$'s an anomaly at finite voltage
(FBA) is apparent.  The position of this FBA is roughly  $\sim g_b E_D$ for
these intermediate values. (see however below). This FBA
is actually built on top of a ZBA. Fig. \ref{fig:gb2.5} shows
the effect of magnetic scattering on the FBA for $g_b = 2.5$.
We see that the FBA  is sensitive to $\gamma$, and is removed
for $\gamma \sim g_b E_D $ (also see, however, below).  The ZBA
 remains after the FBA is destroyed. 

\begin{figure}
\centerline{ 
	\epsfysize=0.4\textwidth \rotate[l]{
	\epsfbox{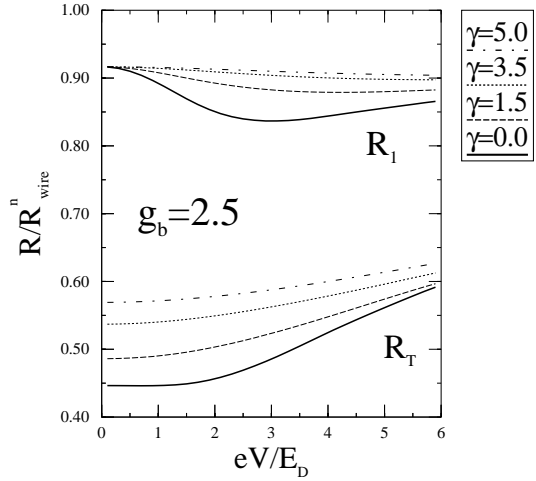} }
}
\caption{
The effective "resistances" $R_1$ and $R_T$ for $g_b=2.5$ for some
values of 
$\gamma$ listed in the legend. $R^n_{wire} \equiv ( G^n_{wire})^{-1} $.
$R_2$ is very small, roughly constant ($ \sim 0.08 R^n_{wire}$)
and is not shown here. }
\label{fig:rsgb2.5}
\end{figure}

To understand these features, it is helpful to consider the `resistances'
defined in eqs (\ref{GT})and (\ref{R1}), 
as shown in Fig. \ref{fig:rsgb2.5} again for $g_b=2.5$. \cite{noteR1}
The FBA results from the dip at finite voltage for $R_1$ 
(more precisely $R_1 + R_2$, the total effective
 resistance for the wire), whereas the
ZBA is from the minimum at zero voltage for $R_T$.  The relatively small
$\gamma$ ( $\sim g_b E_D$ for these intermediate $g_b$)
first remove the feature in $R_1$.  The minimum at $V=0$ for
$R_T$ is somewhat more robust.

The features for $R_1$ and $R_T$ can in turn be related to the
behavior of $\theta_r$ and $\theta_i$. 
(recall eqs(\ref{GT}) and (\ref{R1}). It is also helpful to observe
that a natural length scale $ \sqrt { D \over \epsilon} $
i.e., the coherence length at energy 
$\epsilon$, arises in eq (\ref{bulkt}) )
 Fig. \ref{fig:theta} shows these parameters
again for $g_b = 2.5 $ and for a specific energy $\epsilon = 2.5$.
First consider $\gamma=0$. 
We see that $\theta_i$ is finite within the wire, but small
at the two ends ($\theta_i \approx 0$ at $x=L$ because
$\epsilon << \Delta$).  This $\theta_i$ is generated from the gradients,
as is evident from taking the imaginary part of eq(\ref{bulkt}) :
at finite $\epsilon$ and $\theta_r$ there is a contribution
to a negative curvature of $\theta_i$ in the form of
$ {2 \epsilon \over D}  {\rm sin} \theta_r {\rm cosh} \theta_i$. 
Fig. \ref{fig:thetab} shows the values of 
$\theta_r$ and $\theta_i$ at the immediate
left of the barrier ($x_{b-}$) for two values of $g_b$ at $\gamma=0$.
These values are good indicators of the values of 
$\theta_r$ and $\theta_i$ within the wire if we keep in mind their
typical spatial dependence as in Fig. \ref{fig:theta}.
As the energy increases $\theta_i(x_{b-})$ first increases, then
saturates and
 shows a slight decrease afterwards.  This
gives rise to the shape of $R_1$.
 At larger energies $\theta_r$ decreases faster from the boundary
towards $x=0$, thus
 acquiring a larger slope at the boundary and hence a larger
$[\theta_r]_b$ (as well as a smaller value within the wire which 
slow the increase of $\theta_i$ as just mentioned),
 and gives rise to the increase of $R_T$ as a function of
voltage.  For smaller $g_b$'s the resistance is dominated by
$R_T$, where the angle between the spectral vectors on
the two sides of the barrier are also larger.  Thus
for smaller $g_b$ as a function of voltage $R_T$ increases rapidly with voltage 
and dominates the resistance, and only the ZBA but not the FBA remains.
Thus in short the ZBA is due to the tilting of the spectral vector at $x_{b-}$
at low voltages alway from $\theta_r=0$ due to the proximity effect , and
the FBA is due to the fact that at finite energies 
a spatial gradient increases the
magnitude of the DOSV within the wire.
The presence of a small $\gamma$ suppresses the negative curvature
of $\theta_i$ and thus $\theta_i$ itself, thus destroying
the FBA. 
$\gamma$ also increases the curvature of $\theta_r$ and thus
$[\theta_r]_b$,
which contributes to the suppression of the ZBA. 
(see Fig. \ref{fig:theta} )

An example for the conductance at small  $g_b$ ($0.1$)
is shown in Fig. \ref{fig:gb0.1}. \cite{notewidth}
  It is interesting to note that the curves for different $\gamma$'s
actually cross each other  
at $eV { > \atop \sim} E_D$, i.e., for some
"large" voltage that $G$ actually increases with increasing 
pair-breaking.  This anomalous dependence has also been found
by Volkov and Klapwijk \cite{volkov92b}
for the double barrier structures for large barrier 
resistances.  It is not clear how to understand this physically.

For very large  $g_b$ the resistance of the system
comes almost entirely from the wire, and the FBA is most
prominent.  
The position of the FBA
as well as the $\gamma$ needed to destroy the FBA 
seems to saturate for large $\gamma$ (at $\sim 5 E_D$ and 
$\sim 3 E_D$ respectively):  
see Fig. \ref{fig:gb24} ($g_b=24$). Since increasing $\epsilon$
tends to increase $\theta_i$, for large $\gamma$ the conductance
actually has a minimum at $eV =0$ and increases with increasing voltage.

It is of interest to compare Figs 
\ref{fig:gb},\ref{fig:gb2.5},\ref{fig:gb0.1},\ref{fig:gb24}
 with the results
of Marmorkos et al \cite{marmos93}, who have studied the ZBA and FBA 
 by employing the scattering matrix method.
The qualitative behaviors of the conductance are similar.
  Our FBA
seems to be  located at a similar value of $eV$ and has similar width
\cite{notepeak}.
  However, important differences have to be noted. It is useful first
 to recall that the normal state conductance of the system
is $G^n_{wire} / (1 + g_b^{-1})$ ( e.g, $0.8 G^n_{wire}$ for $g_b=4.0$
and $0.96 G^n_{wire}$ for $g_b=24$).  Hence for $g_b {> \atop \sim} 4$
the conductances at $eV =0$ are almost the normal state values, 
and
 for the voltages shown for these $g_b$'s the
 conductances are always larger than the corresponding
normal state values except for very small voltages.
The conductances at $eV=0$ found
in \cite{marmos93} are smaller than the normal state even for
perfectly transmitting interface, and only overshoot the normal
state value near the FBA for relatively transparent barriers\cite{noteg}.  

Marmorkos et al\cite{marmos93}, and Takane and Ebisawa (TE) \cite{takane93}
had also examined the effect of 
magnetic field $B$ on the conductance at zero  voltage.  The
magnetic field makes the problem intrinsically (at least) two dimensional
in general and we hope to return to that in the future.  However,
since the magnetic field is also a pair-breaker it is of interest
to compare their results with 
Fig. \ref{fig:gb2.5}, \ref{fig:gb0.1} and \ref{fig:gb24}.  It is obvious
from our plots that $G(V=0,\gamma)$ always decreases with increasing $\gamma$.
The relative change in $G$ (i.e. the change measured in
$G(V=0, \gamma) / G(V=0,\gamma=0)$ ) decreases as the barrier resistance
decreases, and vanishes as $g_b \to \infty$
(see Fig. \ref{fig:gb24}, and also \cite{noteR1}) .
In TE \cite{takane93} the behavior of $G(V=0,B)$ as a function of
 increasing magnetic field $B$  
is similar to the one for $G(V=0,\gamma)$ just described.
However in Marmorkos et al although $G(V=0,B)$ still decreases with $B$
for high potential barriers, 
it increases with field for the transparent
interface.  It is unclear why the results of Marmorkos et al and TE are
different. 
% However if we just compare our results with \cite{marmos93}
%and recall that the Usadel equation
% had only included impurity scattering in the self-consistent
%Born approximation,  their interpretation of the suppression of their
%conductivity near $V=0$ as due to enhanced weak-localization is 
%consistent with our results.

For normal metals if $v_f \sim 10^8 cm/sec $ and $ L \sim \mu m$,
then $E_D \sim 0.02 meV$  if $ l \sim 1000 \AA$.  To observe the
FBA the condition $ G^n_b > G^n_{wire}$ roughly requires
the transmission probability (appropriately angular averaged) through 
the barrier to be ${> \atop \sim} 0.1$, which should be easily achievable
for a suitable choice of the metal and the superconductor.
To observe the FBA in the semiconductor-superconductor system
seems to be not so easy.  Even for a choice of semiconductor like
InGaAs where the tunneling through the Schottky barrier is enhanced
by the small effective mass, since the Fermi wavelength on the 
two sides differ significantly most of the excitations are still
reflected.  In order to get   $ G^n_b > G^n_{wire}$ smaller $l$ 
and/or larger $L$ have to be chosen, making $E_D$ much smaller.
One should also keep in mind that most of the recent experiments
on these structures involve semiconductors in the clean limit,
where the present calculation does not apply.

This research is supported by the National Science Foundation through the
Northwestern University Materials Science Center, grant number DMR 91-20521.

\newpage

\begin{figure}
\centerline{ 
	\epsfysize=0.4\textwidth \rotate[l]{
	\epsfbox{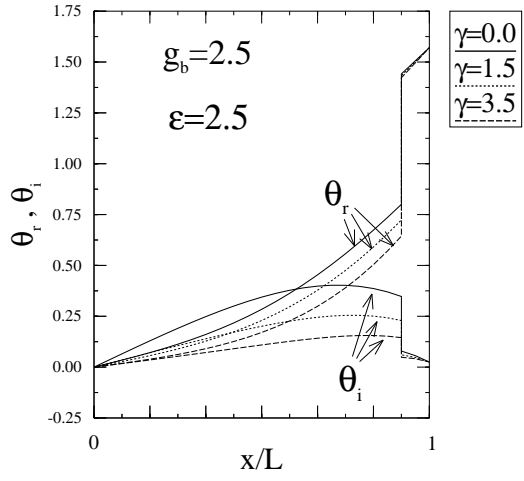} }
}
\caption{
The parameters  $\theta_r$ and $\theta_i$ for $g_b=2.5$ and $\epsilon=2.5$
for some values of 
$\gamma$ listed in the legend. }
\label{fig:theta}
\end{figure}

\begin{figure}
\centerline{ 
	\epsfysize=0.4\textwidth \rotate[l]{
	\epsfbox{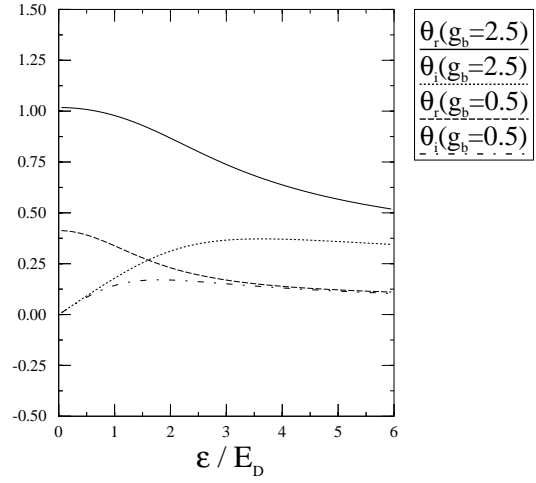} }
}
\caption{
The parameters  $\theta_r$ and $\theta_i$ at $x_{b-}$ 
for $g_b=2.5$ and $g_b=0.5$
as a function of energy.
 $\gamma = 0$.   }
\label{fig:thetab}
\end{figure}

\begin{figure}
\centerline{ 
	\epsfysize=0.4\textwidth \rotate[l]{
	\epsfbox{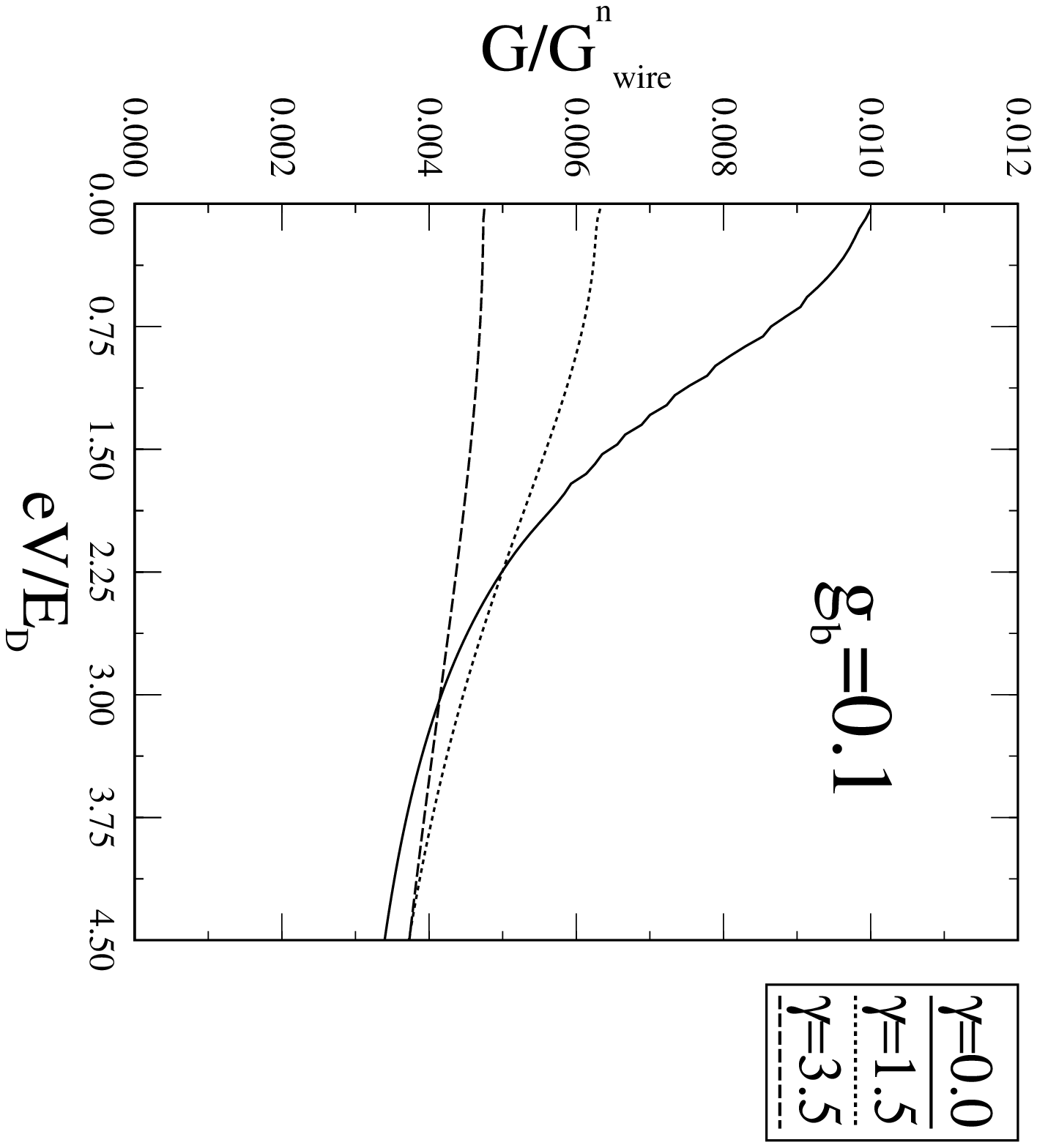} }
}
\caption{
The (normalized) conductance versus voltage at $g_b = 0.1$ for
the values of $\gamma$ listed in the legend. }
\label{fig:gb0.1}
\end{figure}

\begin{figure}
\centerline{ 
	\epsfysize=0.4\textwidth \rotate[l]{
	\epsfbox{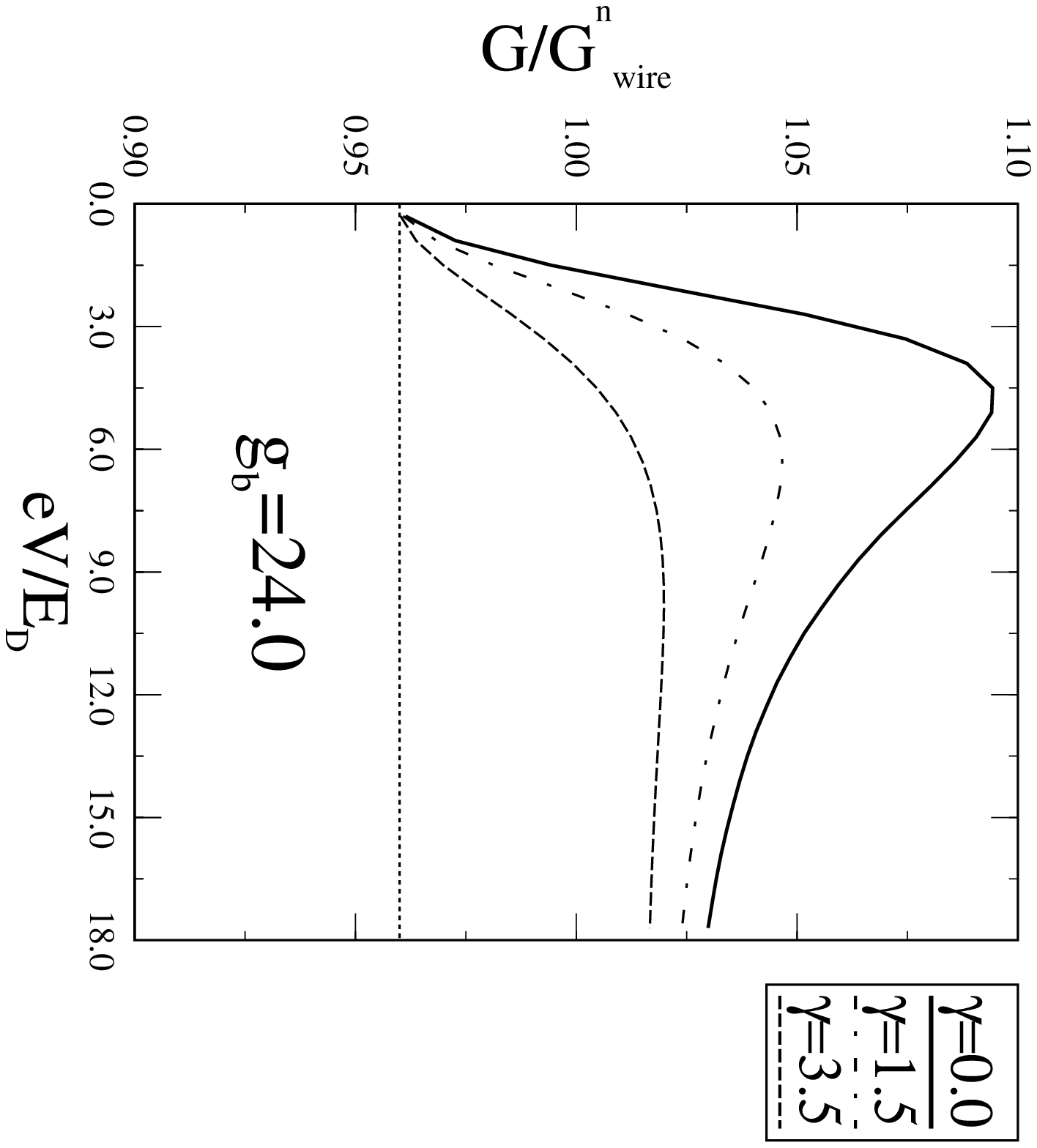} }
}
\caption{
The (normalized) conductance versus voltage at $g_b = 24$ for
the values of $\gamma$ listed in the legend. 
The horizontal dotted line is the normal state conductance of the system.}
\label{fig:gb24}
\end{figure}


\begin{thebibliography}{9}
\bibitem{noteR1}  Notice that  $R_1(V=0)$ is simply
$ 0.92 R^n_{wire}$ (recall $x_b = 0.92 L$)
 and is independent of $\gamma$ since
 we always have $\theta_i(x) = 0 $ at $\epsilon = 0 $
(see eq(\ref{bulkt}). 
However, typically $R_T > R^n_b$ ( $ = g_b^{-1} R^n_{wire} = 0.4  R^n_{wire} $
here), which is  due to the ${\rm cos} [\theta_r]_b $ factor.
\bibitem{notepeak} See their Fig. 2a.  Note that since their calculation
is for a two dimensional lattice their
 $E_c \equiv { \pi \over 2} { v_f l \over L}$ should correspond to
$ \sim 3.14 E_D$ here).
\bibitem{notewidth}  Notice that for $\gamma=0$ the width of the ZBA here 
is still $\sim E_D$, 
not $ \sim g_b^2 ( 1 + g_b )^2 / 4 $ as the calculation of 
ref \cite {volkov92b} for  double barrier systems would seem to suggest
 if we put one of their barrier resistance to zero.
\end{thebibliography}
\end{document}